# Use of bremsstrahlung radiation to identify hidden weak β⁻ sources: feasibility and possible use in radio-guided surgery

**D. Carlotti,**[a,b,c] **F. Collamati,**[a] **R. Faccini,**[a,e,1] **P. Fresch,**[a] **F. Iacoangeli,**[a] **C. Mancini-Terracciano,**[a] **M. Marafini,**[a,f] **R. Mirabelli,**[a,d] **L. Recchia,**[a] **A. Russomando,**[g,h] **E. Solfaroli Camillocci,**[a,e] **M. Toppi,**[i] **G. Traini,**[a,d] **V. Bocci,**[a,1]

[a] *INFN Sezione di Roma,*
  *Roma, Italy*

[b] *Scuola di Specializzazione di Fisica Medica,*
 *Sapienza Universitá di Roma, Roma, Italy*

[c] *U.O.C. Fisica Sanitaria,*
  *Az. Osp. San Camillo-Forlanini, Roma, Italy*

[d] *Dipartimento di Scienze di Base e Applicate per Ingegneria,*
  *Sapienza Università di Roma, Roma, Italy*

[e] *Dipartimento di Fisica,*
  *Sapienza Università di Roma, Roma, Italy*

[f] *Museo Storico della Fisica e Centro Studi e Ricerche "E. Fermi",*
  *Roma, Italy*

[g] *Center for Life Nano Science@Sapienza,*
  *Istituto Italiano di Tecnologia, Roma, Italy*

[h] *Centro Científico Tecnológico de Valparaíso-CCTVal,*
 *Universidad Técnica Federico Santa María, Chile*

[i] *Laboratori Nazionali di Frascati INFN,*
  *Frascati, Italy*

  *E-mail:* Valerio.Bocci@roma1.infn.it
      Riccardo.Faccini@roma1.infn.it

---

[1] Corresponding author.

Abstract: The recent interest in β⁻ radionuclides for radio-guided surgery derives from the feature of the β radiation to release energy in few millimeters of tissue. Such feature can be used to locate residual tumors with a probe located in its immediate vicinity, determining the resection margins with an accuracy of millimeters. The drawback of this technique is that it does not allow to identify tumors hidden in more than few mm of tissue. Conversely, the bremsstrahlung X-rays emitted by the interaction of the β⁻ radiation with the nuclei of the tissue are relatively penetrating.

To complement the β⁻ probes, we have therefore developed a detector based on cadmium telluride, an X-ray detector with a high quantum efficiency working at room temperature. We measured the secondary emission of bremsstrahlung photons in a target of Polymethylmethacrylate (PMMA) with a density similar to living tissue. The results show that this device allows to detect a 1 ml residual or lymph-node with an activity of 1 kBq hidden under a layer of 10 mm of PMMA with a 3:1 signal to noise, i.e. with a five sigma discrimination in less than 5 s.

**Contents**

**1 Introduction**

Primary goal of surgical oncology is the identification and complete resection of the tumor and its residues. Radio-guided surgery (RGS) is a surgical adjunct to achieve this goal, by detecting radiation selectively deployed to the tumor by radio-tracers. While RGS techniques so far mainly focused on γ or β⁺ Radionuclides [1–3], the β emitting radionuclides release their energy within few mm and thus the use of an intra-operative probe sensitive only to the beta radiation allows to delineate the margins with a precision of a few millimetres [4].

The limitation of the technique is that β detectors identifies the tumor tissue only in the immediate vicinity. To overcome this limit, we propose to exploit the X-ray photons produced by bremsstrahlung radiation since these photons can penetrate more than β rays.

Since the applications of the radio-guided surgery with β⁻ decays studied so far make use of DOTATOC-$^{90}Y$ as tracer [5], in this paper we evaluate the activity of $^{90}Y$ needed to have a significant signal with a Cadmium Telluride (CdTe) device behind few mm of equivalent tissue [6]. To this aim, a $^{90}Sr$ is used to evaluate the response of the detector to bremsstrahlung photons, while a $^{90}Y$ source is used to to study its performance in a realistic clinical environment. The developed detector still requires to be adapted to medical applications and it will eventually have to be integrated with the β⁻ probe and/or in an imaging device.

## 2 Material and Methods

### 2.1 Choice of the X-ray sensor

The detector was optimized to maximize the efficiency for bremsstrahlung radiation above 20 keV, to achieve an energy resolution of few keV, to operate at room temperature, and to have a size compatible with the tip of an intra-operative device.

Among solid-state detectors, germanium (Ge) and silicon (Si) detectors have low band gaps which lead to the generation of high dark currents, reported in Tab.1. Ge detectors need therefore to be operated at cryogenic temperatures and Si detector have a low efficiency in the range of interest. CdTe detector have instead a good efficiency in the range of interest with only 1 mm of thickness [7].

**Table 1**. Physical properties of semiconductors at T=25 ºC. [7]

| Material | Si | Ge | CdTe | CdZnTe |
|---|---|---|---|---|
| **Atomic number** | 14 | 32 | 48, 52 | 48, 30, 52 |
| **Density (g/cm$^3$)** | 2.33 | 5.33 | 6.20 | 5.78 |
| **Band gap (eV)** | 1.12 | 0.67 | 1.44 | 1.57 |
| **Pair production (eV)** | 3.62 | 2.96 | 4.43 | 4.6 |
| **Resistivity ($\Omega$ cm)** | $10^4$ | 50 | $10^9$ | $10^{10}$ |
| **$\mu_e \tau_e$ (cm$^2$/V)** | > 1 | > 1 | $10^{-3}$ | $10^{-5} - 10^{-2}$ |
| **$\mu_h \tau_h$ (cm$^2$/V)** | ~ 1 | > 1 | $10^{-4}$ | $10^{-5}$ |

For this study we then opted for a 1 mm thick CdTe sensor with a band gap of 1.44 eV at a working temperature of 25 °C. The detector works with a ohmic contact and it collects the electrons. In recent years some companies began to commercially produce these types of detectors and our choice fell for the EURORAD sensor S.5.5.1.U in TO39 package with a crystal dimension of 5 x 5 x 1 mm$^3$. Its electrical parameters at 25 °C and at 50 V of bias voltage are: resistivity $\rho = 8 * 10^8$ $\Omega$ cm (R=$\rho$ l/S ), leakage current $I_c$ =10 nA, capacity smaller than 5 pF. The charge transport properties for both electrons ($\mu_e \tau_e = 1.5 * 10^{-3}$ cm$^2$/V ) and holes ($\mu_h \tau_h = 1 * 10^{-4}$ cm$^2$/V ) are good. The expected energy resolution of > 5 keV (FWHM) is appropriate [8]

### 2.2 Front-End electronics

A CdTe detector has an electron/hole pair production energy of 4.43 eV, which means that for a typical photon energy of 59.5 keV (like the $^{241}Am$) the detector produces 2.2 fC of charge. To read the electric charge signal of the detector, a system composed of a Charge Sensitive Preamplifier (CSP) followed by a pulse shaping circuit, to transmit the signal in a band-limited channel, and a 12 bits digital oscilloscope (Teledyne Lecroy HDO6104) with histogram capability as Multi Channel Analyzer and 1GHz of bandwidth. The main block of the CSP circuit was a CREMAT [9] hybrid circuit, the polarization network [10] being designed to accomplish the EURORAD S.5.5.1.U leakage current, thus maximizing the signal to electronic noise ratio. We have chosen a CREMAT module amplifier with an amplification of 1400 mV/pC corresponding to a 3 mV output signal for the $^{241}$Am peak. The shaper module reduces the noise by filtering a particular range of frequencies and it therefore reduces the fall time of the pulse signal. The shaping time was chosen to minimize the equivalent noise charge (ENC) estimated as [11].

$$Q_n^2 = 12 \frac{e^2}{nA\,ps} I_d \tau + 6*10^5 \frac{e^2 k\Omega}{ns} \frac{\tau}{R_b} + 3.6*10^4 \frac{e^2 ns}{(pF)^2 (nV)^2/Hz} \frac{e^2}{\tau} C^2 \quad (2.1)$$

where $Q_n$ is in equivalent electrons, $\tau$ is the shaping time, $I_d$ correspond at the detector bias current plus amplifier input current, $C$ is the total input capacitance, $R_b$ is input shunt resistance and $e_n$ is the equivalent input noise voltage spectral density. Fig.1 shows the corresponding dependence on the shaping time. To minimize the ENC, the optimal shaping time turned out to be 450 ns.

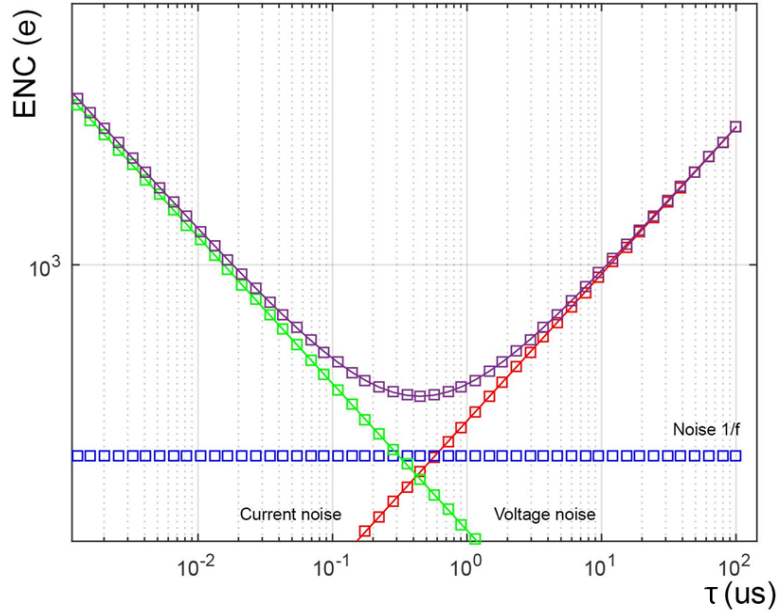

**Figure 1**. Equivalent Noise Charge (see Eq.2.1) as a function of the shaping time ($\tau$).

Eq.2.1 takes into account three noise components, the voltage, the current and the 1/f noise, but it doesn't consider any external noise. In the experimental practice we saw, instead, the presence of various sources of noise with a frequency of approximately 100 kHz coming from switching power supplies ( power charger, computer power supply . . . ). To suppress them we chose a shaping time of 250 ns (see Fig.2 ).

### 2.3 Mechanical Setup

To be confident about the geometry of the experimental setup, a dedicated mechanical setup was developed. We implemented a mechanism to regulate the distance between the source and the detector and to allow the insertion of layers of plastic or metal in a controlled way. This was achieved with a system of micro-metric screws as shown in Fig.3.

### 2.4 Calibration and optimization

The instrument was calibrated with Cs-137, Ba-133, Am-241 and Co-57 sources. Tab.2 lists their main γ lines and β⁻ end points.

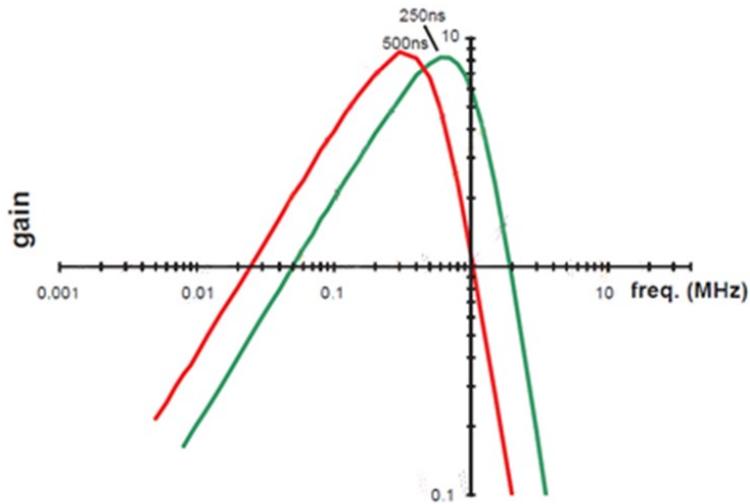

**Figure 2**. CR-200 Bandpass filtering properties [9]: gain as a function of frequency for several shaping times.

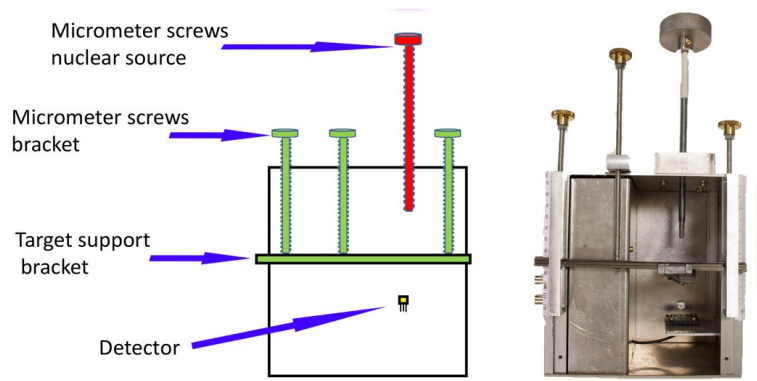

**Figure 3**. The experimental mechanichal structure, designed to regulate the distance of the source to detector: three micrometric screws (green) are used to maintain the flatness of the target support bracket, while a fourth one allows to regulate the distance between the source and the detector (red).

The measured spectra (in mV of output voltage) are shown in Fig.4. To estimate the calibration constant of our instrument, we fitted each peak to a gaussian with an offset, as shown in Fig. 5. Fig.6 shows the dependence of the measured signals on the photon energy from which a calibration constant of α= 1.66 ± 0.07 (mV/keV) is extracted.

We have also studied the dependence of the FWHM on the energy (Fig.6, on the right) The FWHM depends marginally on the energy of the photon and it is on average 7.10 ± 0.04 keV.

## 2.5 Experimental Set-Up with $^{90}Sr$ source

Fig.7 shows the experimental setup used for the measurements with $^{90}Sr$ source: between the detector and the source, a block of PMMA of dimensions 1 x 8 x 8 cm$^3$, can be interposed without changing the relative position of the two. This arrangement has the aim to shield the detector from electrons produced by the nuclear source and at the same time generate bremstrahalung photons.

**Table 2**. Main γ lines and β⁻ end points of the sources used to calibrate the detector in the range of interest. *I* are the relative intensities. [12]

| Source | Attivity (μCi) | Half Life | Radiation | E (keV) | I (%) |
|---|---|---|---|---|---|
| $^{241}Am$ | 1.1 | 432.2 y | γ | 26.3 | 2.4 |
| | | | | 36.2 | 0.13 |
| | | | | 43.4 | 0.07 |
| | | | | 59.5 | 35.9 |
| $^{137}Cs$ | 1.11 | 30.07 y | X | 31.8 | 2 |
| | | | | 32.2 | 3.76 |
| $^{137}Cs$ | 1.11 | 30.07 y | β | 514 | 94.4 |
| $^{57}Co$ | 1.38 | 271.79 d | γ | 14.4 | 9.16 |
| | | | | 122.06 | 85.6 |
| | | | | 136.5 | 10.7 |
| $^{57}Co$ | 1.38 | 271.79 d | β | 700 | 99.8 |
| $^{133}Ba$ | 1.03 | 10.51 y | γ | 79.6 | 2.62 |
| | | | | 81 | 34 |
| $^{133}Ba$ | 1.03 | 10.51 y | X | 30.6 | 35 |
| | | | | 31 | 64.5 |
| $^{133}Ba$ | 1.03 | 10.51 y | β | 80.39 | 86 |
| $^{90}Sr$ | 0.08 | 28.79 y | β⁻ | 546 | 100 |
| $^{90}Y$ | 0.33 | 64.00 h | β⁻ | 2280.1 | 99.99 |

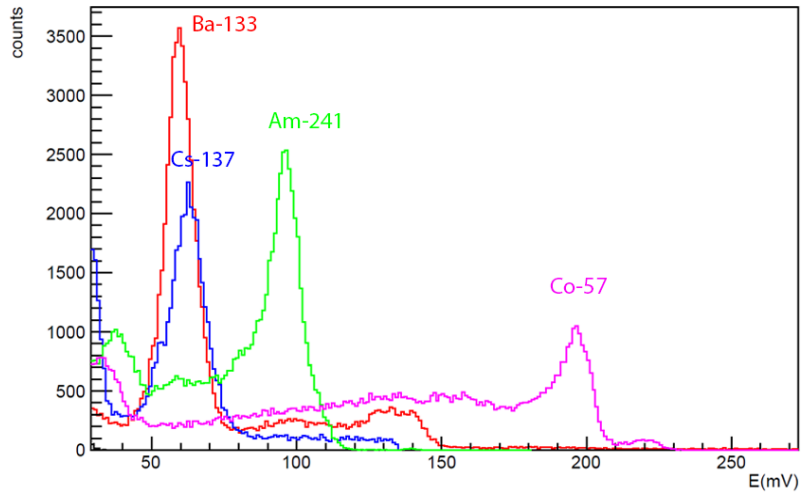

**Figure 4**. Spectrum of the measured energies (in mV of output voltage) obtained with the calibration sources.

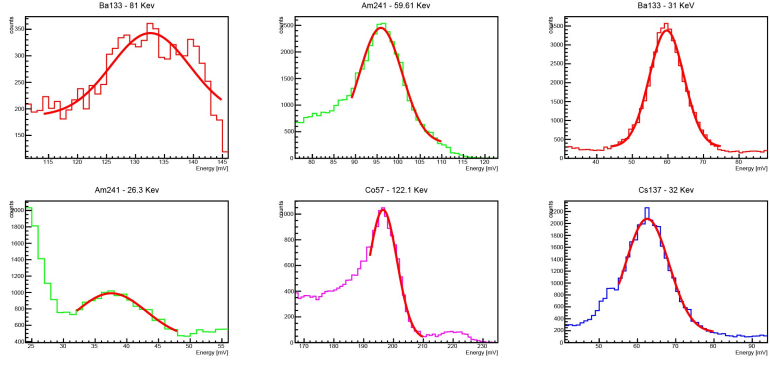

**Figure 5**. Results of the fits to the spectra obtained with the calibration sources.

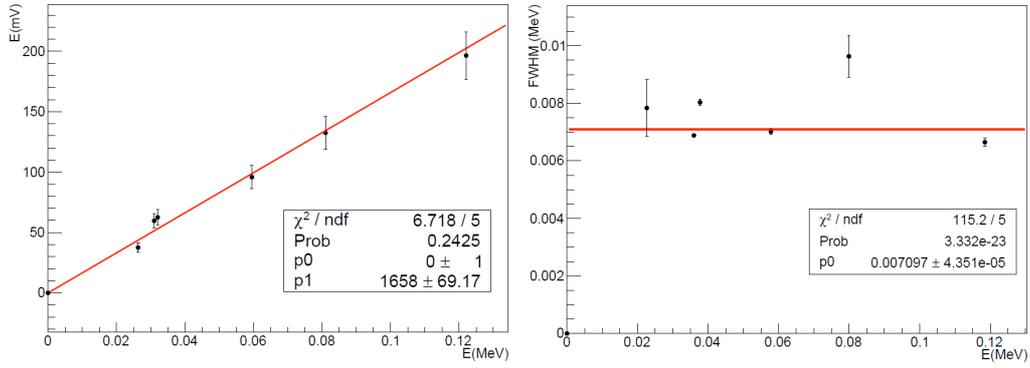

**Figure 6**. Calibration curve (left): detected signal vs photon energy. Dependence of the FWHM of the peaks on the photon energy (right).

We have used an extended $^{90}Sr$ source, distributed on a cylindrical substrate of 1.5 cm diameter and 0.4 cm height with a 0.08 µCi activity over 2π of solid angle. The $^{90}Sr$ decays $\beta^-$ into $^{90}Y$ with an end-point energy of 546 keV, and the latter in turn decays $\beta^-$ with a 2.2 MeV end-point. The mean penetration range in PMMA of the two spectra is ≈1 and ≈4 mm respectively. Due to the $^{90}Sr$ long half-life, $^{90}Sr$ and $^{90}Y$ are in secular equilibrium. To ensure that the 10 mm of PMMA will shield the detector from $^{90}Sr$ beta rays, we utilized an ArduSiPM [13] particle detector characterized by a SiPM detector optically coupled to a plastic scintillator BC-408 with an efficiency which is high for beta radiation and low for X-rays. The lack of signal after shielding confirms that eventual signals on the CdTe detector originate from bremsstrahlung radiation.

## 2.6 Experimental Set-Up with a $^{90}Y$ source

In the case of the $^{90}Y$ source the experimental set-up is analogue to the one for $^{90}Sr$ except for the geometry of the source, since the $^{90}Y$ it is diluted in 1 ml of saline solution, contained within a glass vial. The initial activity of the source was 0.33 µCi and three different measurements were done in the course of 5 days.

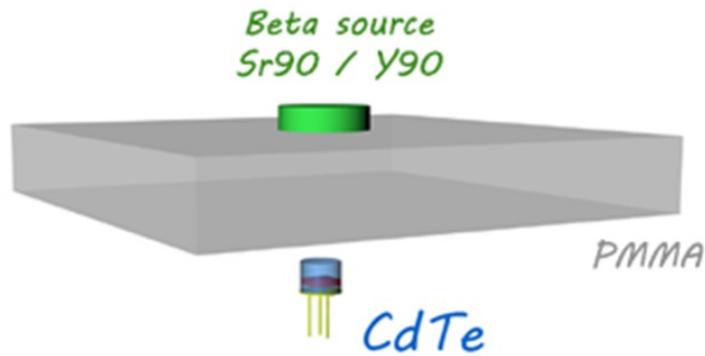

**Figure 7**. Experimental set-up the CdTe detector (blue); the PMMA target (grey); the distributed nuclear source (green).

## 3 Results

We report the spectra obtained from our measurement apparatus, produced by beta sources of $^{90}Sr$ and $^{90}Y$, shielded by a 10 mm thick block of PMMA.

### 3.1 Measurements with $^{90}Sr$

Fig. 8 shows the spectrum acquired with the 0.08 μCi $^{90}Sr$ source compared with a noise spectrum acquired with the same setup but without the source. In both cases the acquisition time was 1 hour. The noise can be ascribed to the environmental radioactivity of our laboratory and to the leakage current of our detector: these two components depend on the ambient temperature and the environmental radioactivity.

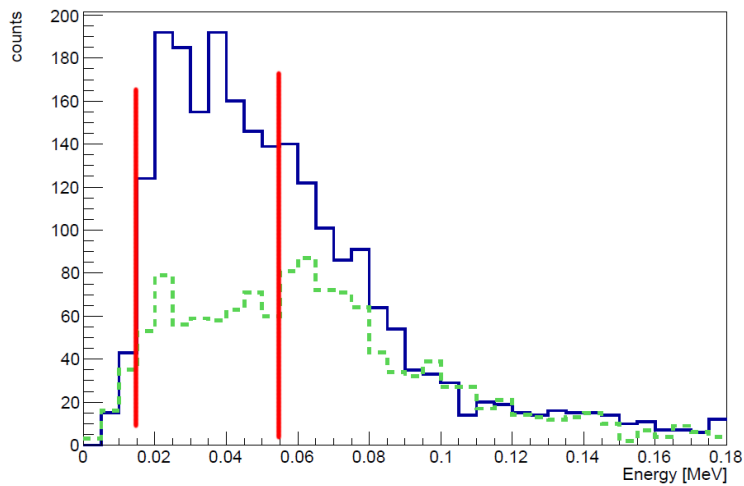

**Figure 8**. Measured energy spectrum in absence (green dashed histogram) and in presence (full blue histogram) of the $^{90}Sr$.

When operating the device, the quantity of interest will be the counting rate integrated in a given energy interval. To determine the optimal interval we computed, as a function of the upper edge, a quantity proportional to the signal significance,

$$Q = S/\sqrt{(B)} \tag{3.1}$$

(where S is the signal rate and B the background rate).

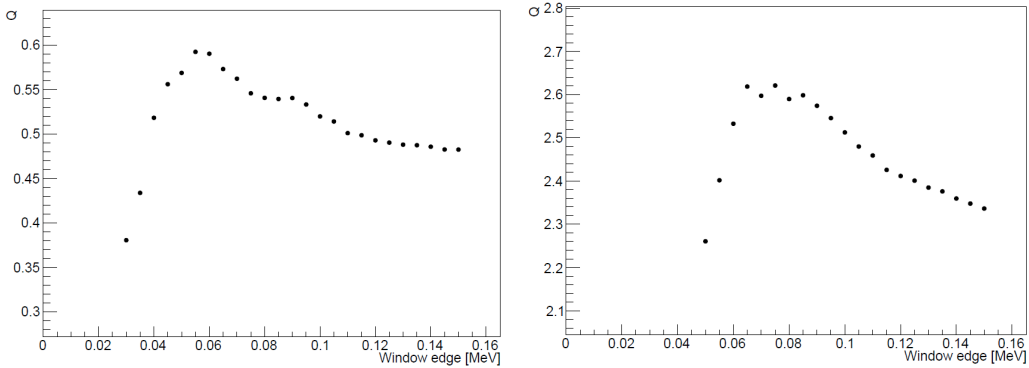

**Figure 9**. $S/\sqrt{B}$ as a function of the upper edge of the energy interval in the case of a $^{90}Sr$ (left) or $^{90}Y$ (right) source.

The results (see Fig. 9) show that the optimal window is 15-55 keV where the observed rates are 0.36±0.03 and 0.14±0.06 cps with and without source, respectively.

## 3.2 Measurements with $^{90}Y$

The $^{90}Sr$ β spectrum is softer than the one from $^{90}Y$, that is supposed to be used in RGS. We have therefore repeated the measurements with the $^{90}Y$ source, obtaining the results in Fig.10, which compares the spectra taken with the $^{90}Y$ source and the noise with an acquisition time of 1 hour when the activity was 0.33 µCi . As expected the spectrum with the source is harder than in the case of $^{90}Sr$, corresponding to a larger optimal window: 20-75 keV (see Fig. 9). The observed rates are 1.98 ± 0.03 and 0.37±0.02 cps with and without source respectively, when the activity was 0.33 µCi. It is to be noted that, as already mentioned, the noise has instabilities that are still under study.

Finally, we have verified the linearity with the activity by repeating the measurements while the $^{90}Y$ was decaying (its half-life is 64 hours). Fig.11 shows the background subtracted rates for signals in the 20-75 keV window as a function of the source activity. The behaviour is clearly linear with a slope k= 5.2 ± 0.3 (cps/µCi).

In order to compare with the expected rates and spectra, a full Monte Carlo (MC) simulation of the tests with $^{90}Y$ was set up with GEANT4 [14]. The full geometry of the setup was included and the particle traced to the detector, where the deposited energy was computed. A particle is considered as "detected" if it deposits energy in the detector.

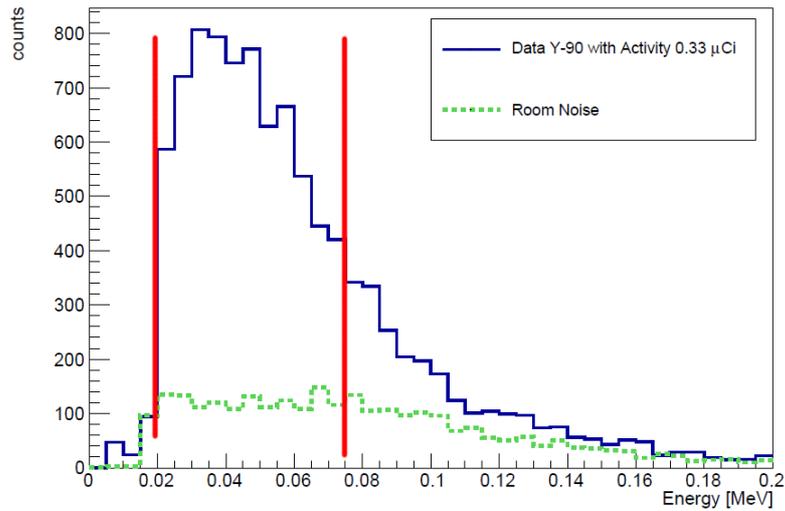

**Figure 10**. Measured energy spectrum in absence (green dashed histogram) and in presence (full blue histogram) of the $^{90}Y$.

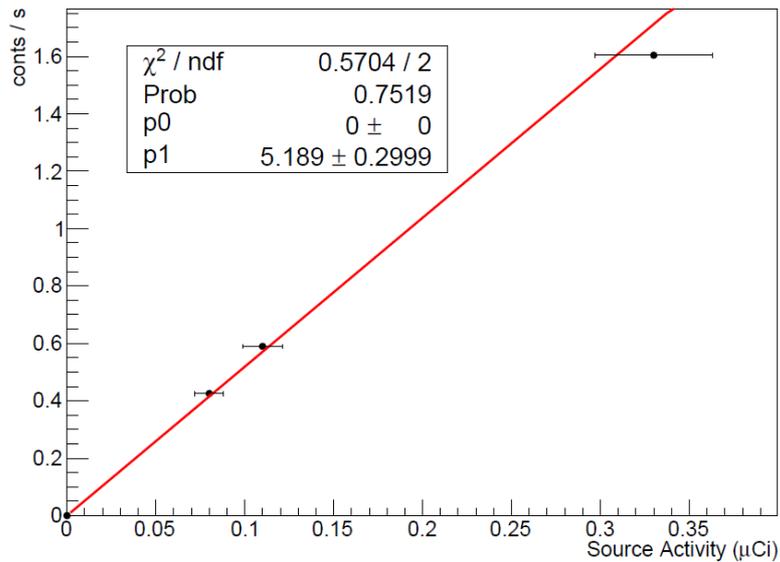

**Figure 11**. Background subtracted signal counts per second with the $^{90}Y$ source as a function of the source activity. The calibration curve is superimposed.

Fig.12 shows the distribution of the energy expected in the CdTe detector with no additional efficiency correction compared to the source spectrum subtracted of the noise. The spectra show a reasonable agreement in shape, while the MC rate is 2% higher than the data. Given the uncertainties in the geometry of the setup, in particular the distribution of the $^{90}Y$ inside the vial and the position of the detector, the differences could be due either to an energy dependent inefficiency or to such uncertainties.

Given the good agreement between data and simulation, to study the dependence of the signal observed as a function of the material interposed between the source and the detector, we simulated the expected signal with different thicknesses of the PMMA shield, normalized to the same number of generated events (Fig. 13). The same figure reports the integrated yields as a function of the thickness. It shows that for PMMA shields thinner than 5 mm there is a component of β⁻ radiation traversing the shield, while between 5 and 10 mm there is no significant reduction of the bremsstrahlung radiation. This is compatible with the expected projected range of the electrons and attenuation length of the X-rays.

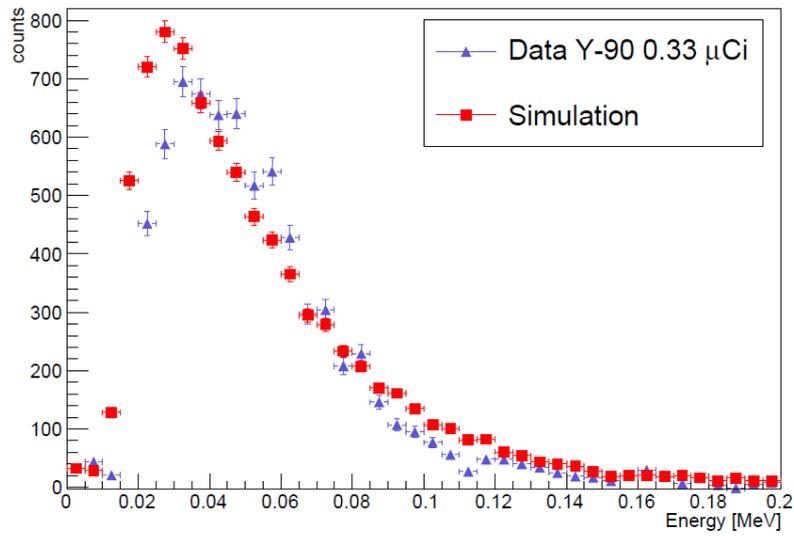

**Figure 12**. MC (red squares) and data (blue triangles) background subtracted energy spectra of the $^{90}Y$ source.

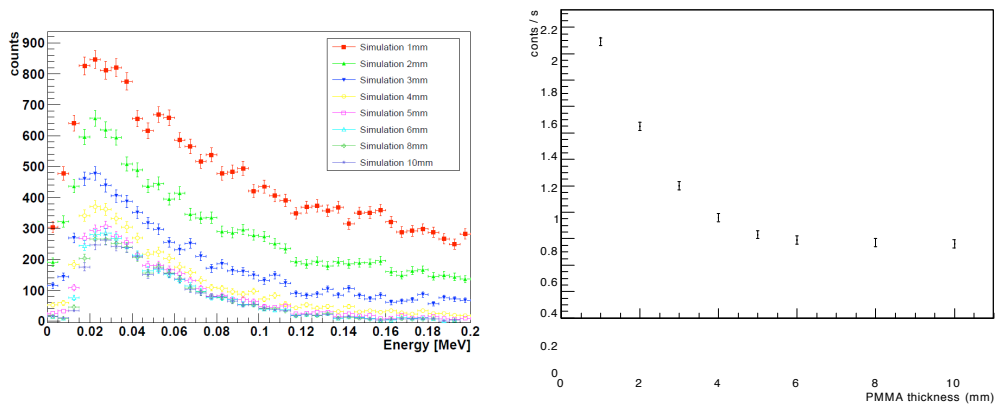

**Figure 13**. Simulated energy spectrum (left) and MC estimate of the counting rates (right) as a function of the PMMA thickness with a source activity of 0.11 μCi.

## 4 Discussion

The typical inter-operative use of bremsstrahlung detection from a $^{90}Y$ source is the search for captating lymph-nodes or small lesions under the surface. The lower limit on the total activity of the searched source, which is used as benchmark, would be 1 kBq (10 kBq/ml in 0.1 ml).

The measurements presented in this paper, scaled according to a Monte Carlo Simulation, show that under these conditions the measured rate of signal would be 0.14 cps, with 0.37 cps of background.

The time needed to distinguish the signal from the noise is then estimated by requiring, under the assumption of Poisson distribution of the counts

$$N_{tot} > N_{noise} + n_\sigma \sqrt{N_{noise}} \qquad (4.1)$$

Where $N_{tot}$ is the number of counts in detector during the integration time, $N_{noise}$ is corresponding noise contribution and $n_\sigma$ is the number of $\sigma$ of discrimination required. In the above mentioned conditions, a $3\sigma$ discrimination would require 170 s of integration, a $5\sigma$, 472 s.

Such an integration time would be compatible with a surgical intervention if an imaging device were designed to give an image over the whole surgical field. On the other side a probe-like device would be sensitive to 1 ml lesions, for which only 4.7 s are required for a $5\sigma$ and 1.7 s for a $3\sigma$ discrimination. Such relatively large remnants could be of interest if hidden under healthy tissue. Further developments in the level of noise of the efficiency would correspondingly impact the sensitivity either in terms of integration time or size of the smallest detectable lesion.

Finally, it is to be noted that the current setup is missing the lateral shielding that would allow directionality of the detector, another feature required in a probe for radio-guided surgery.

## 5 Conclusions

We have investigated the potentialities of a CdTe detector to detect bremsstrahlung from tumor residuals emitting $\beta^-$ radiation from $^{90}Y$ decays hidden under healthy tissues. Assuming a 10 mm equivalent thickness between the source and the detector we have estimated that a 1 ml lesion with 0.27 µCi/ml specific activity would yield a signal of 1.4 cps over a noise of 0.37 cps. The corresponding time to have a $5\sigma$ detection is therefore 4.7 s. A 0.1 ml lesion with the same specific activity would require 7 minutes and 52 seconds to be detected in an imaging device. We have also verified that the rates and spectra are reproducible with a full simulation of the experimental setup and studied the behaviour according to the PMMA thickness.

## Acknowledgments

The authors would like to thank Dr. R. Lunadei and Dr. G. Chiodi for helping us to build the measure equipment.